\documentclass{article}
\usepackage{amsmath}
\usepackage{amsfonts}
\usepackage{graphicx}
\usepackage{amssymb}
\newcommand{\nd}{\noindent}
\newcommand{\be}{\begin{equation}}
\newcommand{\ee}{\end{equation}}
\newcommand{\ben}{\begin{eqnarray}}
\newcommand{\een}{\end{eqnarray}}

 \newtheorem{thm*}{Theorem}

\begin{document}

\title{Correlated Gaussian Systems exhibiting additive
Power-Law  Entropies}
\author{C.\ Vignat\\E.E.C.S., University of Michigan, U.S.A. \\ and L.P.M., E.P.F.L., Lausanne, Switzerland \\\\ A. Plastino\\ Facultad de C. Exactas,
National University La Plata and \\ Argentina's National Research
Council (CONICET) \\ C. C. 727, 1900 La Plata, Argentina\\\\ A. R.
Plastino\\ Physics Department, University of Pretoria,\\ Pretoria
0002, South Africa}

\maketitle

\maketitle \nd {\bf Abstract} \vskip 1mm \nd We show, on purely
statistical grounds and without appeal to any physical model, that
a power-law $q-$entropy $S_q$, with $0<q<1$, can be {\it
extensive}. More specifically, if the components $X_i$ of a vector
$X \in \mathbb{R}^N$ are distributed according  to a  Gaussian
probability distribution $f$, the  associated entropy $S_q(X)$
exhibits the extensivity property for special types of
correlations among the $X_i$. We also characterize this kind of
correlation.\vspace{0.2 cm} \nd PACS: 05.30.-d, 05.30.Jp

\nd KEYWORDS: Extensivity, Gaussian systems, Power-law entropies.
\section{Introduction}

\nd The Boltzmann-Gibbs logarithmic expression

\be \label{BGE} S_{BG}\{p(i)\} = -\sum_i\, p(i)\,\ln{p(i)}, \ee
where one sums over  microstates labelled by $i$ constitutes a
cornerstone of our present understanding of Nature. If one knows a
priori the expectation value $U=\langle H \rangle$ of the
pertinent Hamiltonian, $S_{BG}$ is maximized by   Gibbs'
probability distribution (PD) for the canonical ensemble
\cite{pathria,reif,lavenda,katz}, usually referred to as the
equilibrium Boltzmann-Gibbs (BG)  distribution

\be \label{gibbs}  p_G(i) =\frac{\exp{(-\beta E_i)}}{Z_{BG}}, \ee
with $E_i$ the  energy of the the microstate $i$, $\beta=1/k_B T$
the inverse temperature, $k_B$ Boltzmann's constant, and
$Z_{BG}$ the partition function. This PD can safely be regarded as
the most notorious and renowned PD in the field of statistical
mechanics.

\nd In the last 15 years this PD has found a counterpart in the
guise of power-law distributions with which an ``alternative"
thermostatistics (usually referred to by the acronym ``NEXT") can
be built that constitutes nowadays not only a very active field
but also, for many people, a new paradigm for statistical
mechanics, with applications
 to several scientific disciplines \cite{gellmann,euro,lissia,fromgibbs}.
 Power-law distributions are certainly ubiquitous in physics (critical phenomena are
just  a conspicuous example \cite{goldenfeld}). It is argued
\cite{euro} that, in the same manner that  the BG-thermostatistics
addresses thermal equilibrium states on the basis of Boltzmann's
molecular chaos hypothesis ({\it Stosszahl-ansatz}), the
alternative thermostatistics  deals with phenomena in natural,
artificial, and social systems that do not accommodate with such a
simplifying hypothesis. However, NEXT
 uses the whole BG theoretical machinery, as for instance, the
maximum entropy principle (MaxEnt)  \cite{katz}.

\section{The extensivity question}

\nd The power law  entropy $S_q$, with
nonextensivity index $q \in \mathbb{R}$
\cite{gellmann}

\be \label{dino}  S_q =  \frac{1}{q-1}\,\left(1-\sum_i
p_i^{q}\right); \,\,\,S_1=S_{BG}, \ee
 is a nonextensive
information measure: if $A$ and $B$ are independent systems then a priori
$S_q(A+B) \ne S_q(A) + S_q(B).$ This assertion has come into
question quite recently \cite{euro}. There exist composite systems
for which the correlation among its components is of such a nature
that $S_{BG}$ becomes {\it nonextensive}, while $S_q$ becomes {\it
extensive} \cite{euro,arp,gell,horsch}. These references deal with
exceedingly interesting examples of this new facet of
$q-$thermostatistics. In particular, Refs. \cite{gell,horsch} need
the special concept of {\it asymptotic scale invariance}. Ref.
\cite{arp} considers both a classical and a quantum example in the
thermodynamic limit. In the present communication we address a
somewhat more general situation in the sense that we do not have
any particular model in mind but a whole {\it class} of systems:
Gaussian ones, of enormous importance in several areas of
scientific endeavor.

\section{Our main result}

We will not need appeal neither to the {\it Stosszahl-ansatz} nor
to {\it asymptotic scale invariance}. Our arguments are of a
purely statistical nature, which can be regarded as giving them
more generality. We will concern ourselves in what follows with
the continuous instance in $\mathbb{R}^N$. Let the $N$ components
of vector $X \in \mathbb{R}^N$ be distributed according to a
probability distribution $f$ and write Tsallis' entropy
$S_{q}\left(X\right)$ as \be \label{dino2}
S_{q}\left(X\right)=\frac{1}{q-1}\left(1-\int f^{q}\right).\ee

\vskip 3mm

\nd Our main result is easily stated and reads as follows. Let $f$
be of a Gaussian character. Then:
\begin{thm*}

If $\,0\le q\le1$ and $N\in \mathbb{N}$ then there exists a positive definite matrix $K$ and an $N-$variate Gaussian
vector $X$ with covariance matrix $K$ such 
that $X$ {\sf verifies the extensivity condition}

\be \label{result}
S_{q}\left(X\right)=\sum_{i=1}^{N}S_{q}\left(X_{i}\right).\ee

\end{thm*}

\subsection{Proof of the theorem}

A centered Gaussian random variable with dimension $N$ and
covariance matrix $K$ has a probability distribution function
(PDF)

\be \label{GPDF} f_{X}\left(X\right)=\frac{1}{\vert2\pi
K\vert^{1/2}}\exp\left(-\frac{1}{2}X^{T}K^{-1}X\right), \ee and
thus  its associated $S_q$ entropy is easily seen to write

\be \label{dino1}  S_{q}\left(X\right)=\frac{1-\vert2\pi
K\vert^{\frac{1-q}{2}}q^{-N/2}}{q-1}. \ee One may assume without
loss of generality that each component $X_{i}$ of $X$ has unit
variance so that its associated  $q-$sub-entropy reads

\be \label{dino3}
S_{q}\left(X_{i}\right)=\frac{1-\left(2\pi\right)^{\frac{1-q}{2}}q^{-1/2}}{q-1}.\ee

\nd Now, the condition for extensivity adopts the appearance

\be \label{extens}
S_{q}\left(X\right)=\sum_{i=1}^{N}S_{q}\left(X_{i}\right).\ee
Alternatively one can write

\be \label{extens1} 1-\vert2\pi
K\vert^{\frac{1-q}{2}}q^{-N/2}=N\left(1-\left(2\pi\right)^{\frac{1-q}{2}}q^{-1/2}\right),\ee
or,  equivalently,

\begin{equation}   \vert
K\vert^{\frac{1-q}{2}}= N\left(2\pi\right)^{\left(1-N\right)
\frac{1-q}{2}}q^{\frac{N-1}{2}}-
\left(N-1\right)q^{N/2}\left(2\pi\right)^{N\frac{q-1}{2}}.
\label{eq:detK}\end{equation}

\nd Notice that the right-hand side above is positive, since
the function $f_{N}\left(q\right)$

\begin{equation}
\label{funcion}
f_{N}\left(q\right)=\frac{N}{N-1}\,\,
\frac{\left(2\pi\right)^{\frac{1-q}{2}}}{{\sqrt{q}}},
\,\,\,\,\,\,\,0\le q\le1
\end{equation}
decreases with $q$ in this interval, verifies
$f_{N}\left(1\right)=\frac{N}{N-1}>1$ and thus $f_{N}(q) > 1$ on $[0,1]$.
Thus, equation
(\ref{eq:detK}) can be solved as\begin{equation} \label{eq:det}
\vert K\vert=\left(N\left(2\pi\right)^{\left(1-N\right) \,
\frac{1-q}{2}}q^{\frac{N-1}{2}}-
\left(N-1\right)q^{N/2}\left(2\pi\right)^{N\frac{q-1}{2}}\right)^{\frac{2}{1-q}}.\end{equation}

\nd It only remains now to find a matrix $K$ with unit diagonal
entries that satisfies the  condition (\ref{eq:det}). This entails
that  one has to choose $\frac{N\left(N-1\right)}{2}$ correlation
coefficients \be \label{choose}
K_{i,j}=E\left[X_{i}X_{j}\right]=\langle
X_{i}X_{j}\rangle;\,\,\, 1\le i<j\le N.\ee
In the two following subsections, we provide two examples of such matrix.

\subsection{Example: all uncorrelated components except for two of them}

An obvious choice for the off-diagonal elements of $K$ is\[
K_{i,j}=\begin{cases}
\sigma & i=2,\,j=1  \,\, \textrm{ or }   \,\, i=1,\,j=2\\
0 & \text{else} \,\, (i \ne j) \end{cases}\] so that $\vert K\vert=1-\sigma^{2}$
where $\sigma$ should verify\[
\begin{cases}
\sigma^{2}=1-\left(N\left(2\pi\right)^{\left(1-N\right)\frac{1-q}{2}}q^{\frac{N-1}{2}}-\left(N-1\right)q^{N/2}\left(2\pi\right)^{N\frac{q-1}{2}}\right)^{\frac{2}{1-q}}\\
0<\sigma<1\end{cases}\] which has a unique solution since function
\ben & g_{N}:\left[0,1\right]\rightarrow\left[0,1\right]\cr \cr &
g_N(q)= 1-\left(N\left(2\pi\right)^{\left(1-N\right)
\frac{1-q}{2}}q^{\frac{N-1}{2}}-\left(N-1\right)q^{N/2}
\left(2\pi\right)^{N\frac{q-1}{2}}\right)^{\frac{2}{1-q}},\een
 is decreasing with $q$, verifies $g_{N}(0)=1$ and $g_{N}(1)=0$ and thus is one to one. In Fig.1 below, the
  function $g_{N}\left(q\right)$
is plotted versus $q$ for values $N=2, 5, 10$ and $20$ from bottom
to top.


\nd It is apparent that,  for large $N-$values, this curve
approaches a  constant (equal to unity) in the interval
$\left[0,1\right[$, while it vanishes for q=1. This means that,
{\sf for a fixed value of $q$, a more and more intense correlation
degree becomes necessary between $X_{1}$ and $X_{2}$ as $N$ grows
so as to ensure extensivity for the system at hand}. The same
conclusion is reached in Ref. \cite{arp} for their models.

\subsection{Example: all equally correlated components}

Another obvious choice is to select a correlation matrix of the
form\[ K_{i,j}=\begin{cases}
\sigma & i\ne j\\
1 & \text{else}\end{cases}\] so that each pair of distinct
components has the same degree of correlation. In such
scenario  we have\[ \vert
K\vert=\left(1-\sigma\right)^{N-1}\left(1+(N-1)\sigma\right)\] which

is, for a given $N$, a decreasing function $g(\sigma)$  of
$\sigma$ such that  $g(0)=1$ and $g(1)=0$.
Thus, for given values of $N$ and $q$, there exists a
unique value of $\sigma$ such that condition (\ref{eq:det}) holds.
In Fig. 2 these values of $\sigma$ are plotted as a function of
$q$ for $N=2,\,3$ and $5$ respectively.


\nd Identical  conclusions to those of the preceding subsection
are arrived at here, except that  a smaller  correlation degree is
now needed to  reach extensivity, because the correlation is in
this instance  spread over a larger number of components than
previously.

\section{Conclusion}

We have shown that, for {\it any} Gaussian distributed vector in
$\mathbb{R}^N$, Tsallis' entropy becomes {\it extensive} for $q
\in [0,1]$ if an adequate type of correlation among its components
exists. This correlation imposes  special conditions on the
covariance matrix that apply only to its determinant and can
always be determined. Moreover, we provided a constructive proof
that explicitly yields the covariance matrix that fulfills the
desired purpose. We did not need appealing  neither to the {\it
Stosszahl-ansatz} nor to {\it asymptotic scale invariance}. Our
arguments being of a purely statistical nature, they can be fairly said to considerably generalize the results described in \cite{euro}.


\vspace{1 cm}

\begin{thebibliography}{9}

\bibitem{pathria}
 Gibbs J W  1948 {\it Elementary principles in statistical mechanics
in Collected Works} (New Haven: Yale University Press; R. B.
Lindsay R B and Margenau H 1957 {\it Foundations of physics\/}
 (NY: Dover).

\bibitem{reif} F.~ Reif, {\it Statistical and thermal physics}
(McGraw-Hill, NY, 1965).

\bibitem{lavenda} B. H. Lavenda,
{\it Statistical Physics}
 (Wiley, New York, 1991);
   B. H. Lavenda, {\it Thermodynamics of Extremes} (Albion, West Sussex,  1995).





\bibitem{katz} E. T. Jaynes, Phys. Rev. {\bf 106}  (1957) 620; {\bf
108} (1957) 171;  {\it Papers on probability, statistics and
statistical physics}, edited by R. D. Rosenkrantz (Reidel,
Dordrecht, Boston, 1987);  A.~Katz, {\it Principles of Statistical
Mechanics, The Information Theory Approach} (Freeman and Co., San
Francisco, 1967).


\bibitem{gellmann}  M.~Gell-Mann and C.~Tsallis, Eds.
{\it Nonextensive Entropy: Interdisciplinary applications} (Oxford
University Press, Oxford, 2004); V.~Latora, A.~Rapisarda, and C.~
Tsallis, {\it Physica A} {\bf 305} (2002) 129 (and references
therein); S. Abe and Y. Okamoto, Eds. {\it Nonextensive
statistical mechanics and its applications} (Springer Verlag,
Berlin, 2001);  A. R. Plastino and A. Plastino, Phys. Lett. A {\bf
177}  (1993) 177.

\bibitem{euro} Special Issue, Europhysics-news, {\bf 36} (2005).

\bibitem{lissia}  Kaniadakis G,  Lissia M, and  Rapisarda A., Eds., 2002
{\it Nonextensive statistical mechanics and physical
applications}, Physica A (Special) {\bf 305}, and references
therein.


\bibitem{fromgibbs} A.~ Plastino, A. R. ~ Plastino, {\it Phys. Lett.
A} {\bf 193} (1994) 251.




\bibitem{goldenfeld} N.~ Goldenfeld, {\it Lectures on phase transitions and the
renormalization group} (Addison-Wesley, NY, 1992).


\bibitem{arp}  C. Zander and A.R. Plastino, Physica A (2005) in
Press.

\bibitem{gell} C. Tsallis, M. Gell-Mann, anf Y. Sato.
cond-mat/0502274, also presented by C. Tsallis at the
International Conference on {\it Complexity and Non-Extensivity:
New Trends in Statistical Mechanics}, that took place at the
Yukawa Institute, Kyoto on 14-18 March 2005.



\bibitem{horsch} J. Marsch and S. Earl, Phys. Lett. A (2005) in
Press.



\end{thebibliography}
\end{document}